\documentclass[aps,preprint,amsmath,amssymb,amsfonts,nofootinbib]{revtex4}
\usepackage{epsfig}
\usepackage{graphicx}
\usepackage{dcolumn}
\usepackage{bm}
\usepackage{amsthm}
\usepackage{amsmath}
\usepackage{color}
\usepackage{verbatim}

\newcommand{\beq}{\begin{equation}}
\newcommand{\eeq}{\end{equation}}
\newcommand{\bea}{\begin{eqnarray}}
\newcommand{\eea}{\end{eqnarray}}

\begin{document}

\begin{flushright} 
TAUP-2979/13
\end{flushright}

\title{Bulk Viscosity in Holographic Lifshitz Hydrodynamics}

\author{Carlos Hoyos, Bom Soo Kim and Yaron Oz}
\affiliation{Raymond and Beverly Sackler School of
Physics and Astronomy, Tel-Aviv University, Tel-Aviv 69978, Israel\\
E-mail: choyos,bskim,yaronoz@post.tau.ac.il
}

\begin{abstract}
We compute the bulk viscosity in holographic models dual to theories with Lifshitz scaling and/or hyperscaling violation, using a generalization of the bulk viscosity formula derived in arXiv:1103.1657 from the null focusing equation. We find that only a class of models with massive vector fields are truly Lifshitz scale invariant, and have a vanishing bulk viscosity. For other holographic models with scalars and/or massless vector fields we find a universal formula in terms of the dynamical exponent and the hyperscaling violation exponent.
\end{abstract}

\maketitle


\tableofcontents

\section{Introduction and summary}

The equation of state of Conformal field theories  (CFTs) at finite temperature follows from the
tracelessness of the stress-energy tensor
\begin{equation}
\varepsilon=d p \ ,
\end{equation}  
where $\varepsilon$ is the energy density, $p$ is the pressure and $d$ is the number of spatial dimensions. This linear
relation between the energy density and the pressure takes a generalized form in theories with a dynamical exponent $z$ and a hyperscaling violation exponent $\theta$, it reads
\begin{equation}
z\varepsilon=(d-\theta) p \ .
\end{equation}  
Theories with different values of $z$ and $\theta$ (including CFTs if $z=(d-\theta)/d$) may coincidentally have the same equation of state. Thus, more information is needed in order to determine whether a theory is truly scale invariant. 
In the hydrodynamic description of a CFT at finite temperature, the transport coefficients are constrained by the underlying symmetries of the theory \cite{Baier:2007ix}. A well known example appears already at the first dissipative order, conformal invariance implies that the trace of the energy-momentum tensor should vanish, which forces the bulk viscosity to be zero $\zeta=0$.

A natural question is whether there are any constraints on transport coefficients in theories with some scale symmetry (but not conformal symmetry). Such constraints can provide, among other things, a clear-cut way to distinguish between theories with the same equation of state. 
For Lifshitz theories with a dynamical exponent $z$ invariant under the transformation $x^i \to \lambda x^i, \ \ t \to \lambda^z t$, the equation of state follows from the Ward identity for the energy-momentum tensor that we derived at the ideal level in \cite{Hoyos:2013eza,Hoyos:2013qna} 
\begin{equation}
zT^{\mu\nu}u_\mu u_\nu-P^{\mu\nu}T_{\mu\nu}=0 \  ,
\label{WI}
\end{equation}
where $u^\mu$ is the velocity of the fluid $u^\mu u_\mu=-1$. When $z=1$ it reduces to the identity in conformal field theories $T^\mu_\mu=0$. At the first dissipative order, the expression (\ref{WI}) will be non-zero if the bulk viscosity is non-zero, thus signalling a breaking of Lifshitz scale invariance. So, as for the conformal case, one would reach the conclusion that $\zeta=0$ for Lishitz theories. However, in contrast to the conformal case, the Ward identity in Lifshitz theories depends on the velocity. This means, in particular,  that the generator of scale transformations (as well as the other symmetry generators) depends on the velocity. Thus,
whether (\ref{WI}) holds beyond the ideal order is far from clear. In this paper we will present evidence that the na\"{\i}ve Ward identity is still valid, by computing the bulk viscosity in gravitational models that are holographic duals to theories with Lifshitz scaling.

Holographic models with Lifshitz scaling have attracted much attention in recent years, partly because of their potential application to condensed matter physics \cite{Kachru:2008yh,Taylor:2008tg} (see also \cite{Gath:2012pg} for various models of interest to us at zero temperature). Lifshitz solutions to Einstein equations can be obtained when massive vector fields, scalar and massless vector fields together, or a combination of all of them are present. There is also the possibility of using higher form fields to construct Lifshitz solutions, but we will not treat those here, except when they are equivalent to one of the previous cases.

For theories with a holographic dual, there is a simple way to compute the bulk viscosity from the null focusing equation at the horizon \cite{Eling:2011ms,Eling:2011ct} based on the framework of  \cite{Eling:2009pb,Eling:2009sj,Eling:2010hu}. This successfully captures the bulk viscosity for various models, such as the hydrodynamics of non-conformal theories  \cite{Mas:2007ng,Benincasa:2006ei,Kanitscheider:2009as}, perturbations of the N = 4 supersymmetric Yang-Mills theory \cite{Benincasa:2005iv,Buchel:2008uu,Buchel:2010gd,Yarom:2009mw} and holographic models of QCD \cite{Gubser:2008sz,Gursoy:2009kk}. Here we apply the null focusing equation
technique to  (mainly) theories with Lifshitz asymptotics.

For models with massive vectors only (including non-dynamical scalars) \cite{Balasubramanian:2009rx,Korovin:2013nha,Bertoldi:2009vn,Bertoldi:2009dt}, we find that the bulk viscosity indeed vanishes
\begin{equation}
\zeta_{V} =0 \ .
\end{equation}
As we will explain in detail in \S \ref{SecMassive}, the massive vector field in all these models is dual to a marginal operator in the Lifshitz theory, which is only possible for a fixed mass $m^2=zd$. For all other cases, the Lifshitz scaling symmetry is broken either explicitly in the metric if there is hyperscaling violation, or by other background gauge fields (massive scalars or vectors with $m^2\neq zd$). Interestingly, we find that the ratio between the bulk and shear viscosities in these models takes the universal form
\begin{equation} \label{BVmassless} 
\frac{\zeta_{\phi}}{\eta}= -2 \frac{\theta  }{d(d-\theta)} +2\frac{z-1 }{d-\theta} \ .
\end{equation} 
This reminds of the universal value of the shear over entropy density ratio \cite{Kovtun:2004de}. The formula (\ref{BVmassless}) is valid for both neutral fluids such as \cite{Charmousis:2009xr,Pang:2009ad,Gubser:2009qt,Goldstein:2009cv,Goldstein:2010aw,Charmousis:2010zz,Iizuka:2011hg,Gouteraux:2011ce,Cadoni:2011nq} and charged fluids such as \cite{Tarrio:2011de,Alishahiha:2012qu}, although
in general the value of the bulk viscosity will depend on the charges. We provide an explicit example that demonstrates this
in \S \ref{secRunning}, using \cite{Charmousis:2010zz}. When $z=1$ we recover from (\ref{BVmassless}) the value of the bulk viscosity for non-conformal branes \cite{Mas:2007ng}, which can be understood via the compactification of a higher dimensional conformal theory \cite{Kanitscheider:2009as} 
\begin{align}
\frac{\zeta_{z=1,\theta}}{\eta} = -2 \left[  c_s^2 (z=1, \theta) - c_s^2 (z=1, \theta=0)  \right] 
= -2 \left[  \frac{1}{d-\theta} - \frac{1}{d} \right] \ ,
\end{align}
where $c_s$ is speed of sound. Thus, presumably the dependence on $\theta$ can be explained in general using the compactification of a higher dimensional theory with broken Lifshitz symmetry. 
When $z > 1$, there is an additional contribution that can be written as  
\begin{align}
\frac{\zeta_{z,\theta}}{\eta} = 2 \left[  c_s^2 (z, \theta) - c_s^2 (z=1, \theta)  \right] 
= 2 \left[  \frac{z}{d-\theta} - \frac{1}{d-\theta} \right] \;. 
\end{align} 
At weak coupling, the result \eqref{BVmassless} is likely to have a different functional dependences on the difference of the speed of sounds. 

The outline of the paper is as follows. In \S \ref{secFormula}, we derive the general formula for the bulk viscosity using the horizon focusing equation for models with scalars and vector fields. In \S \ref{SecMassive} we study models with massive vector fields. In \S \ref{secScalarNeutral} and \S \ref{secScalarCharged} we study other models of neutral and charged fluids respectively. In \S \ref{secRunning} we study the dependence of the bulk viscosity with the charge in a particular model. We speculate about the implications of our results for the physical properties of quantum critical points in \S \ref{secConclusion}. We provide the relevant Einstein and Maxwell equations in appendix, \S \ref{EinsteinEQ}.

\section{General formula for the bulk viscosity} \label{secFormula}

A general class of holographic models are Lifshitz solutions to Einstein gravity coupled to scalars and Abelian vector fields [$ A, B = (r, \mu) = (t, x_1, \cdots, x_d)$].  The bulk action reads
\begin{align}\label{holoaction}
  S \!=\! \int\! d^{d+2}x \sqrt{-g}\! \left[\! R \!-\! \sum_{j} \!\left(\! \frac{Z_j (\phi)}{4} F_{j AB} F_j^{AB} \!+\! \frac{1}{2} m_j^2 V_{j A} V_j^A \!\right)\! - \! \sum_{i} \!\left(\! \frac{1}{2}(\partial\phi_i)^2 \!-\! V(\phi_i) \! \right) \!  \right] \;,  
\end{align}
where $V_{j A}$ are massive or massless (for $m_j=0 $) vector fields, $F_{j AB}$ their field strengths, and  $\phi_i$ are the scalar fields. $Z_i (\phi)$ parametrize the couplings between scalar fields and vector fields. In Lifshitz solutions rotational invariance is not broken, so only the $V_{jr}$ and $V_{jt}$ components of the vector fields can be non-zero. 

For $(d+1)$-dimensional systems with dynamical exponent $z$ and hyperscaling violation exponent $\theta$, the metric of the gravity dual is
\begin{equation}
 d s^2 =\rho^{-2 +2\theta/d} \left(- \rho^{-2(z-1)} f(\rho) d t^2 
		+ \sum_{i=1}^d (d x^i)^2 +\frac{d \rho^2}{f(\rho)}\right),
\end{equation}
where $f(\rho)$ is the black body factor ($f(\rho)=1$ at zero temperature). Although the majority of analytic black hole solutions are written in this form, we find it more useful for our analysis to use `domain wall' coordinates 
\begin{align} \label{ExplicitMetric}
  &ds^2 = e^{2A(r)}\left[- e^{g(r)} d t^2 + \sum_{i=1}^d (d x^i)^2 \right] + \frac{ dr^2}{e^{g(r)}} \;,  \\
  &e^{A(r)} = r^{\frac{d-\theta}{(z-1)d -\theta}} \;, \quad 
  e^{g(r)} = r^{\frac{2d(z-1)}{(z-1)d -\theta}}  f(r) \;. 
\end{align}
We will call this the `Lifshitz metric' or `Lifshitz solution' for brevity. The scaling symmetries of the field theory appear as transformation properties of the metric in the gravity dual. They map to the geometric transformations
\begin{equation}\label{eq:scaling}
x^i \to \lambda x^i, \ \ t \to \lambda^z t,\ \ r\to \lambda^{-(z-1)+\theta/d} r.
\end{equation}
When the hyperscaling exponent $\theta$ is non-zero, the metric changes by an overall rescaling
\begin{equation}
ds^2\to \lambda^{2\theta/d}ds^2.
\end{equation}

Finite temperature states are dual to black hole geometries. The metric in these cases
takes the same form \eqref{ExplicitMetric}, but $ f(r)$ is a now a function of the radial coordinate that vanishes at the horizon $ f(r_H)=0$ and goes to one at the boundary $f(r\to \infty)=1$. We should point out that this is not the most general possible form of the metric for a black hole solution. There could be sub-leading corrections to the function $A(r)$ (or alternatively to the metric component $g_{rr}$) that depend on the radius of the black hole horizon. For the specific examples we study, the metric is of the form \eqref{ExplicitMetric}. We comment about other cases in \S \ref{SecMassive}.

\subsection{Scalar and vector contributions to bulk viscosity}

Here we provide a simple and clear way to obtain bulk viscosity in the gravitational description
by generalizing the result of \cite{Eling:2011ms}, where the bulk viscosity of a fluid is obtained from the null horizon focusing equation. First, we rewrite the metric as 
\begin{equation}
ds^2=-2 e^{A}u_\mu dx^\mu dr+(e^{2A}P_{\mu\nu}-e^{2A+ g}u_\mu u_\nu)dx^\mu dx^\nu \ ,
\end{equation}
where $u_\mu$ is a time-like unit vector, $\eta^{\mu\nu}u_\mu u_\nu=-1$ and $P_{\mu\nu}=\eta_{\mu\nu}+u_\mu u_\nu$ is the projector in the transverse directions. In addition to the metric, we allow for scalar and vector fields with a non-trivial radial dependence
\begin{equation}
\phi_i(r), \ \ V_j=V_{jt}(r)u_\mu dx^\mu + V_{jr}(r)dr .
\end{equation}
For massless vectors, we can use a gauge $V_{jr}=0$.

The hydrodynamic equations of motion of the fluid in the holographic dual can be obtained by allowing all the fields to depend on the spacetime coordinates and projecting the Einstein equations using the null normal vector $\ell^A$ and evaluating them at the horizon, where
\begin{equation}
\ell^A=\left(\frac{e^{A+g}}{2},u^\mu \right)\, \longrightarrow\, \left(0,u^\mu \right).
\end{equation}
In particular, the projection of the Einstein tensor is proportional to the divergence of the entropy current
\begin{equation}
R_{AB}\ell^A\ell^B =\frac{2\pi T}{s} \partial_\mu (s \ell^\mu)- \sigma^{\mu\nu} \sigma_{\mu\nu}  + \cdots \;,
\end{equation}
where  $T$  is the temperature, $s$ the entropy density, and $ \sigma_{\mu\nu}$ the shear tensor. This equation receives corrections from the energy-momentum tensor of the matter fields $T_{AB}$:
\begin{equation}
\partial_\mu (s \ell^\mu)=\frac{2\eta}{T} \sigma^{\mu\nu} \sigma_{\mu\nu} + \frac{\eta}{T} \mathcal B  + \cdots \;,
\end{equation}	
where we have used $\eta=s/(4\pi)$ and
\begin{align}\label{calB}
\mathcal B  = 2\,T_{AB} \ell^A\ell^B.
\end{align} 

The projection of the energy-momentum tensor receives two types of contributions, one from the kinetic terms of the scalar fields and the other from the masses of the massive vector fields. To leading order in derivatives,
\begin{equation}
2\,T_{AB} \ell^A\ell^B=\sum_i (\ell^A \partial_A\phi_i)^2 + \sum_j m_j^2 (\ell^A V_{jA})^2 
=\frac{\zeta}{\eta}(\partial_\mu u^\mu)^2 \ ,
\end{equation}
where $\zeta$ is the bulk viscosity. The terms proportional to the kinetic term of the vector fields do not contribute to the bulk viscosity and we will ignore them in the following.

The contribution from the scalar fields was computed in \cite{Eling:2011ms}. This is done by converting spacetime derivatives of the scalar into derivatives of entropy and charges 
\begin{align}\label{eq:scalarproj}
\ell^A\partial_A\phi_i=u^\mu\partial_\mu\phi_i^H=\frac{d\phi_i^H}{ds}\partial_\mu s +\frac{d\phi_i^H}{d\rho^a}\partial_\mu \rho^a=-\left( s\frac{d\phi_i^H}{ds}+\rho^a\frac{d\phi_i^H}{d\rho^a}\right)(\partial_\mu u^\mu).
\end{align}
In this expression $\phi_i^H$ are the values of the scalar fields evaluated at the horizon and $\rho^a$ are global conserved charges in the dual field theory (equal to the number of massless vector fields that provide independent charges). In the last equality we used 
the conservation of the entropy current and the charge currents at leading order in derivatives
\begin{equation}
\partial_\mu(s u^\mu)=0, \ \ \partial_\mu(\rho^a u^\mu)=0.
\end{equation}
Using \eqref{eq:scalarproj} one can show immediately that the scalar contribution to the bulk viscosity $\zeta_s$ is 
\begin{align}  \label{ScalarBulk}
	\frac{\zeta_s}{\eta} = \sum_i \left( s \frac{d\phi_i^H}{ds} 
	+ \rho^a \frac{d\phi_i^H}{d\rho^a} \right)^2 \;.
\end{align}

We will now derive the contribution to the bulk viscosity from massive vector fields.
The equations of motion for the massive vector $V_{jA}$ are
\begin{equation} \label{MaxwellEOM}
\partial_A \left( \sqrt{-g} Z_j(\phi) F^{A}_{j\ B} \right) - m^2 \sqrt{-g} V_{jB}=0. 
\end{equation}
Using the equation of motion of the massive vector field, we can rewrite the vector contribution to ${\cal B}$ as
\begin{equation}
{\cal B}_v=\sum_j \frac{1}{m_j^2}\left[ \frac{1}{\sqrt{-g}}\partial_M\left( Z_j(\phi) \sqrt{-g} g^{MN} g^{AB} F_{j NB}\right) g_{AC} \ell^C \right]^2.
\end{equation}
The non-vanishing terms at the horizon inside the bracket are
\begin{equation}\label{terms1}
e^{-(d+1)A}\partial_r(Z_j e^{(d+1)A}g^{rN} g^{AB} F_{j NB})g_{A\mu}u^\mu+e^{-(d+1)A}\partial_\alpha(Z_j e^{(d+1)A}g^{\alpha N} g^{AB} F_{j NB})g_{A\mu}u^\mu \ ,
\end{equation}
and $g_{A\mu}u^\mu=-e^{A}\delta_A^r$ at the horizon since $e^g\to 0$ there. Then, from the first term we get, up to the overall sign that we can drop,
\begin{equation}
e^{-dA}\partial_r\left[ Z_j e^{(d-1)A}\left(u^\mu u^\nu F_{j\mu\nu}+e^{A+g}  u^\mu F_{j \mu r}+e^{A+g}  u^\mu F_{j r\mu}\right) \right].
\end{equation}
The first term vanishes exactly because of the antisymmetry of $G_{j \mu\nu}$, while the other two terms vanish at the horizon $e^g\to 0$.

From the second term in \eqref{terms1} we get
\begin{equation}
e^{-dA}\partial_\alpha\left[ Z_j e^{(d-1)A}\left(e^{-A}P^{\alpha\mu}u^\beta F_{j \mu\beta}+P^{\alpha\mu}e^g F_{j \mu r} +u^\alpha u^\beta F_{j r\beta}\right)\right].
\end{equation}
The first term is higher order in derivatives, and the second vanishes because $F_{j \mu r}\propto u_\mu$. So we are left to leading order with
\begin{equation}
e^{-dA}\partial_\alpha\left[ Z_j e^{(d-1)A}u^\alpha u^\beta F_{j\, r\beta}\right]
=e^{-dA}\partial_\alpha(Z_j e^{(d-1)A}V_t' u^\alpha ) \ ,
\end{equation}
where prime denotes a derivative with respect to $r$.
We will write it as
\begin{equation}
e^{-dA}\partial_\alpha(Z_j e^{(d-1)A}V_{jt}' u^\alpha )=Z_j e^{-A} V_{jt}'\left[ \partial_\alpha u^\alpha + u^\alpha\partial_\alpha\varphi_j+d u^\alpha\partial_\alpha A\right],
\end{equation}
where we have defined
\begin{equation}
\varphi_j=\log(Z_j e^{-A} V_{jt}').
\end{equation}
This choice will be clear below. 
Using again 
\begin{equation}
u^\alpha\partial_\alpha X = -\left(s\frac{dX}{ds}+\rho^a\frac{dX}{d\rho^a}\right)\partial_\alpha u^\alpha,
\end{equation}
we obtain the contribution of the massive vector fields to the bulk viscosity
\begin{equation}
\frac{\zeta_v}{\eta}=\sum_j\frac{e^{2\varphi_j}}{m_j^2}\left[1-d\left(s\frac{dA}{ds}+\rho^a\frac{dA}{d\rho^a}\right)-\left(s\frac{d\varphi_j}{ds}+\rho^a\frac{d\varphi_j}{d\rho^a}\right)\right]^2.
\end{equation}
Unlike the scalar case, we use the equation of motion for the massive vector field to derive the result. The middle expression comes from the determinant of the metric, and $dA$ is nothing but $\log s$. We note that all the examples we find in the literature have a special property that $\varphi$ is a constant, independent of $s$ and $\rho^a$. These two properties have important implications below.

Combining with \eqref{ScalarBulk}, the contributions to the bulk viscosity from the scalars $\phi_i$ and massive gauge fields $ V_j$ are
\begin{align}\label{BulkViscosityTotal}
\frac{\zeta}{\eta} &= \sum_i \left( s \frac{d\phi_i^H}{ds} 	+ \rho^a \frac{d\phi_i^H}{d\rho^a} \right)^2  \nonumber \\
&+ \sum_j \frac{e^{2\varphi_j}}{m_j^2}\left[1-d\left(s\frac{dA}{ds}+\rho^a\frac{dA}{d\rho^a}\right)-\left(s\frac{d\varphi_j}{ds}+\rho^a\frac{d\varphi_j}{d\rho^a}\right)\right]^2.
\end{align}

\section{Examples with unbroken Lifshitz symmetry} \label{SecMassive}

Models with Lifshitz scaling necessarily involve a background vector field\footnote{Or a higher rank dual form, but we will not consider that case here unless it is equivalent to a massive vector.} in order to break Lorentz symmetry. The mass squared of the vector field (in units of the curvature radius) is related to the scaling dimension $\Delta$ of the dual operator through the formula
\begin{equation}
m^2=(\Delta-z)(\Delta-d).
\end{equation}
Even if the metric has Lifshitz scaling, for a general value of the mass the scaling symmetry is broken by the background vector field. There is a special case where the scaling symmetry is {\em not} broken, this happens when the dual vector operator is marginal, i.e. its scaling dimension is
\begin{equation}
\Delta=z+d.
\end{equation}
Or, equivalently, when the mass of the vector field in the bulk is
\begin{equation}
m^2=zd.
\end{equation}
We will show that in this case the bulk viscosity computed holographically indeed vanishes. This suggests that the Ward identity for the energy-momentum tensor that was derived at the ideal level in \cite{Hoyos:2013eza,Hoyos:2013qna} continues to hold at first order in the hydrodynamic expansion
\begin{equation}
zT^{\mu\nu}u_\mu u_\nu-P^{\mu\nu}T_{\mu\nu}\simeq \zeta \partial_\mu u^\mu=0.
\end{equation}
For all other cases the Lifshitz scaling symmetry is broken, either explicitly in the metric if there is hyperscaling violation, or by other background gauge fields (massive scalars or vectors with $m^2\neq zd$). We compute the bulk viscosity in different classes of examples and derive some general formulas for each class.

\subsection{Vanishing bulk viscosity} 

There are only a handful of analytic black hole solutions involving massive vector fields. In \cite{Balasubramanian:2009rx} the action is of the form \eqref{holoaction} with a single vector field and a scalar. A peculiarity is that the scalar is an auxiliary field, there is no kinetic term for it and therefore it does not contribute to the bulk viscosity. 
A second class of examples involve a single massive vector. Analytic solutions were found in \cite{Bertoldi:2009vn} (for a spherical horizon) and also in \cite{Korovin:2013nha}, although in the latter it is only known for the first terms in an expansion in $z-1=\epsilon^2 \ll 1$.

In all these models the metric takes the form \eqref{ExplicitMetric} (for \cite{Korovin:2013nha} this is to leading order in $z-1$). A convenient combination of $tt$ and $rr$ components of the Einstein equation \eqref{AppEE} is
\begin{align}
0 &= d A''  + \frac{1}{2} e^{-2 (A+g)} m^2 V_t^2 \;,
\end{align} 
where $V_t$ is the time component of the massive vector field and $m^2=zd$ its mass. The massive Maxwell's equation \eqref{AppMax} is
\begin{align} \label{MaxwellMassive2}
&0 = - e^{(d-1) A(r) - g(r)} m^2 V_t + \left(e^{(d-1) A(r)} Z(\phi) V_t' \right)'  \;, 
\end{align} 
where $Z(\phi)=e^{-2\phi}$ in the model with the auxiliary scalar \cite{Balasubramanian:2009rx} or $Z(\phi)=1$ otherwise. Combining these two equations we obtain
\begin{align}  \label{EinsteinEqMassive}
& \left(e^{(d-1) A(r)} Z(\phi)  V_t' \right)' = \sqrt{-2 m^2 d A(r)'' e^{2d A(r)}} \;. 
\end{align}
We can use the expression for $A(r)$ in \eqref{ExplicitMetric} to integrate equation \eqref{EinsteinEqMassive}. As a result we find
\begin{align}
& e^{\varphi} = e^{-A(r)} Z(\phi) V_t' = \sqrt{\frac{2m^2 (z-1)}{d}} \;. 
\end{align} 
Note that $\varphi$ is independent of the radial coordinate. Then, the formula for the bulk viscosity \eqref{BulkViscosityTotal} becomes
\begin{align}\label{BulkViscosityTotal2}
\frac{\zeta}{\eta} &= \frac{e^{2\varphi}}{m^2}\left[1-d\left(s\frac{dA}{ds}\right)\right]^2.
\end{align}
Since
\begin{align}
s = \frac{1}{4} e^{dA(r_H)} \; \Rightarrow \; s\frac{dA}{ds} = \frac{1}{d} \;,
\end{align} 
we find that the bulk viscosity is exactly zero in these models
\begin{equation}
\zeta=0.
\end{equation}
For the model in \cite{Korovin:2013nha} we actually know this to be true only to leading order $O(z-1)$, there could be contributions of higher order $O((z-1)^2)$, however in order to compute them one would need to determine the corrections to the vector field, that have not been computed. 

From the derivation above it is unclear whether the result of vanishing bulk viscosity is valid in general solutions with massive vectors. It depends crucially on the form of the metric. In principle the value of the bulk viscosity could change if $e^A$ were a more complicated function of the radial coordinate. We show now that this actually does not happen for the numerical solutions found in \cite{Danielsson:2009gi,Mann:2009yx,Brynjolfsson:2009ct,Bertoldi:2009vn,Bertoldi:2009dt}.%
\footnote{We note that the black brane solutions \cite{Danielsson:2009gi,Mann:2009yx,Brynjolfsson:2009ct} have an additional $B_{\mu\nu}$ field in addition to massless vector field. By dualizing the two form one can show that the model is equivalent to a massive vector field.}  

In order to facilitate the comparison, we start by writing the metric in the notation of \cite{Bertoldi:2009vn,Bertoldi:2009dt}:
\begin{equation}
ds^2=-e^{2{\cal A(\rho)}}dt^2+e^{2{\cal B(\rho)}}dx_i^2+e^{2{\cal C(\rho)}}d\rho^2, \ \ V=e^{{\cal G(\rho)}}dt.
\end{equation}
The translation to domain wall coordinates is straightforward, we can easily identify ${\cal B}=A$ (${\cal B}$ not to be confused with the bulk viscosity expression in the previous section) and 
\begin{equation}
e^g=e^{2{\cal A}-2{\cal B}}, \ \ \frac{dr}{d\rho}=e^{{\cal A}+{\cal C}-{\cal B}}. 
\end{equation}
Then,
\begin{equation}
e^{\varphi}=e^{-A}V_t'=\frac{e^{-{\cal B}}}{\frac{dr}{d\rho}}\partial_\rho\left(e^{{\cal G(\rho)}}\right)=e^{-({\cal A}+{\cal C})}\partial_\rho\left(e^{{\cal G(\rho)}}\right).
\end{equation}
Close to the boundary, the leading order terms of the metric functions are 
\begin{align}
& {\cal A} = \ln (\rho^z)+\cdots,\\
&{\cal B} = \ln(\rho),\\
&{\cal C}=-\ln(\rho)+\cdots,\\
&{\cal G}=\ln(\rho^z)+\cdots.
\end{align}
Note that the expression is exact for ${\cal B}$. 
On the other hand, close to the horizon $\rho=\rho_0$, the solutions take the form
\begin{align}
& {\cal A} = \ln\left(\rho^z(a_0(\rho-\rho_0)^{1/2}+\cdots)\right),\\
&{\cal B} = \ln(\rho),\\
&{\cal C}=\ln\left(\frac{1}{\rho}(c_0(\rho-\rho_0)^{-1/2}+\cdots)\right),\\
&{\cal G}=\ln\left(\sqrt{\frac{2(z-1)}{z}}\rho^z(a_0 g_0(\rho-\rho_0)+\cdots)\right).
\end{align}
Here $a_0$, $c_0$ and $g_0$ have to be determined by matching the solution close to the horizon with the asymptotic boundary solution. Then,
\begin{align}\label{varphinum}
& e^{\varphi(\rho_0)} =\sqrt{\frac{2(z-1)}{z}} \frac{g_0 \rho_0}{c_0}\;. 
\end{align} 

If we do the coordinate transformation
\begin{equation}
\rho=\rho_0 u, \ \ t=\rho_0^z \tau, \ \ x_i=\rho_0 y_i,
\end{equation}
This transformation is a Lifshitz rescaling, so the asymptotic form of the metric {\em and the vector field} do not change 
\begin{align}
& {\cal A} = \ln (u^z)+\cdots,\\
&{\cal B} = \ln(u),\\
&{\cal C}=-\ln(u)+\cdots,\\
&{\cal G}=\ln(u^z)+\cdots.
\end{align}
The solution close to the horizon becomes
\begin{align}
& {\cal A} = \ln\left(u^z(\hat{a}_0(u-1)^{1/2}+\cdots)\right),\\
&{\cal B} = \ln(u),\\
&{\cal C}=\ln\left(\frac{1}{u}(\hat{c}_0(u-1)^{-1/2}+\cdots)\right),\\
&{\cal G}=\ln\left(\sqrt{\frac{2(z-1)}{z}}u^z(\hat{a}_0 \hat{g}_0(u-1)+\cdots)\right).
\end{align}
Where
\begin{equation}
\hat{a}_0=a_0 \rho_0^{1/2}, \ \hat{c}_0=c_0 \rho_0^{-1/2}, \ \hat{g}_0=g_0\rho_0^{1/2}.
\end{equation}
Matching the two solutions determines the values of $\hat{a}_0$, $\hat{c}_0$ and $\hat{g}_0$. Their value is independent of $\rho_0$, since the asymptotic metric and vector functions are independent of $\rho_0$ in the new coordinates. Therefore $\varphi$ in \eqref{varphinum} is independent of $\rho_0$ and the formula for the bulk viscosity becomes \eqref{BulkViscosityTotal2}, which vanishes.

There exists another class of models that involve a massive and a massless vector fields, considered in  \cite{Pang:2009pd,Dehghani:2011tx,Barclay:2012he}. In this class the only known solutions have a fixed charge for a given temperature, and thus it is not possible to vary independently the entropy and the charge density. This prevents us from applying the bulk viscosity formula \eqref{BulkViscosityTotal}.

\section{Examples with broken Lifshitz symmetry}

We study now general examples where the metric has Lifshitz invariance but the scaling symmetry is broken by other fields, either scalar or vector. Typically they have a running scalar, which introduces hyperscaling violation. For some solutions it is possible to avoid the hyperscaling violation. Nevertheless the Lifshitz scaling symmetry is still broken due to the massless vector fields, which affect to the bulk viscosity through the coupling with the scalar. These properties are manifest in Einstein-Maxwell-Dilaton models \cite{Charmousis:2009xr,Pang:2009ad,Gubser:2009qt,Goldstein:2009cv,Goldstein:2010aw,Charmousis:2010zz,Iizuka:2011hg,Gouteraux:2011ce,Cadoni:2011nq}.   We find that the bulk viscosity is non-zero and has a universal expression in terms of the dynamical and hyperscaling violation exponent for these models.

\subsection{Neutral solutions} \label{secScalarNeutral}

We derive the Einstein equations in Appendix \S \ref{EinsteinEQ} from the action \eqref{holoaction} in the absence of massive vector fields $m_j =0$. A combination of $tt$ and $rr$ components of the Einstein equation \eqref{AppEE} is  
\begin{equation}\label{einsteq1}
 \sum_i (\phi_i')^2  = -2 d A''.
\end{equation}
One can compute bulk viscosity using only the general structure of this equation in domain wall coordinates.  

Lifshitz solutions should include at least a massless vector field to be able to break Lorentz invariance in the presence of the scalar field. However, for the solutions studied in \cite{Charmousis:2009xr,Gubser:2009qt,Charmousis:2010zz,Gouteraux:2011ce,Cadoni:2011nq} (\cite{Goldstein:2009cv,Goldstein:2010aw} for $\theta=0$) this does not introduce an additional conserved charge in the dual field theory. 
The reason is that the boundary metric depends on the electric flux. Then, if the dual field theory is in a space with fixed geometry, the electric flux in the bulk is not allowed to change and there is no associated thermodynamic variable. This means that in spite of having massless vector fields in the bulk, the dual fluid dynamics is neutral. 

Then, from \eqref{BulkViscosityTotal}, 
\begin{align}   \label{BulkViscosity}
	\frac{\zeta}{\eta} &= \sum_i \left( s \frac{d\phi_i^H}{ds} 
	+ \rho^a \frac{d\phi_i^H}{d\rho^a} \right)^2 =\left( s \frac{d\phi^H}{ds} \right)^2 
	= \left( s \left(\frac{ds}{dr_H} \right)^{-1} \frac{d\phi^H}{dr_H} \right)^2 \nonumber \\
		&= -2 \frac{A''(r)}{d A'(r)^2} \bigg|_{r=r_H} = - 2 \frac{\theta  }{d(d-\theta)} +2\frac{z-1 }{d-\theta} \;.
	\end{align}
where we have used $s=\frac{1}{4} e^{dA(r_H)} $ and \eqref{einsteq1}. We have split the bulk viscosity in two terms, the contribution from hyperscaling violation, proportional to $\theta$, and a contribution from the Lifshitz scaling proportional to $z-1$. The latter seems to have the imprint of hyperscaling violation as can be seen in the denominator. 
The general form of the result \eqref{BulkViscosity} comes from Einstein-Maxwell-Dilaton models of  \cite{Charmousis:2009xr,Gubser:2009qt,Charmousis:2010zz,Gouteraux:2011ce}. This result is independent of the details of the potential $V(\phi) $ we choose \cite{Goldstein:2009cv,Gouteraux:2011ce}, signifying its universal features within this class of models with asymptotic Lifshitz symmetry. 

The result \eqref{BulkViscosity} includes the known results of the non-conformal branes \cite{Mas:2007ng,Kanitscheider:2009as} as special cases when $z=1$. One can explicitly check this as 
\begin{align}
\frac{\zeta}{\eta} = 2 \left( \frac{1}{d} - c_s^2\right) \;, \quad c_s^2 = \frac{5-d}{9-d} = \frac{1}{d-\theta} \;.
\end{align}
Where $c_s^2$ is the speed of sound in the non-conformal theory.

\subsection{Charged solutions} \label{secScalarCharged}

There are models with several massless vector fields and a single massless scalar \cite{Tarrio:2011de,Alishahiha:2012qu} with $\theta=0$. For $N$ massless vector fields there are $N-1$ independent charges $Q_i, i = 1, \cdots, N-1$. The electric flux for the remaining vector field, as in the neutral case, cannot be varied if the boundary metric is fixed.
In these models the value of the scalar field at the horizon turns out to be simply $\phi(r_H)=r_H^\alpha$, where $\alpha$ is a model-dependent exponent. The entropy is $s=e^{d A(r_H)}/4$. This implies that the variation of $\phi(r_H)$ at {\em fixed entropy} vanishes. Then, the derivation of the bulk viscosity follows through in the same way as for the neutral case and we recover \eqref{BulkViscosity}. 

While charge is not an independent thermodynamic variable for a single scalar case \cite{Charmousis:2010zz}, it can be the case for more complicated matter contents such as two scalar fields with massless vector \cite{Gouteraux:2011qh}. The authors of \cite{Gouteraux:2011qh} provide a bulk viscosity formula for charged hydrodynamics from the compactification of a $2\sigma$-dimensional conformal field theory on a $2\sigma-d-1$ torus. The compactified theory is in general charged. For the neutral case the bulk over shear viscosity ratio reduces to the case of non-conformal branes (\eqref{BulkViscosityTotal} with $z=1$). For the general charged case, the difference with \eqref{BulkViscosityTotal} can be parametrized in terms of the difference between the speed of sound of the charged $c_s$ and neutral $c_s^2=1/(d-\theta)$ case.
\begin{align}
\frac{\zeta_c}{\eta} = -2 \frac{\theta}{d(d-\theta)}+ 2
\frac{d-\theta}{d-1-\theta}\left(\frac{1}{(d-\theta)^2}-c_s^4\right)\;.
\end{align} 
It would be interesting to see if a similar formula applies for more general backgrounds. 

It is worth noting that solutions that interpolate between Anti-deSitter at the boundary and Lifshitz at the horizon are charged even for a single gauge field \cite{Goldstein:2009cv,Cadoni:2009xm,Bertoldi:2010ca,Bertoldi:2011zr,Chemissany:2011mb,Cadoni:2011nq,Berglund:2011cp,Ogawa:2011bz,Huijse:2011ef}  and therefore we expect that the result for the bulk viscosity changes in view of the result given below in \S \ref{secRunning}. We also note that the bulk viscosity of the IR Lifshitz fixed point (with spatial anisotropy) with AdS$_5$ asymptotics has been computed previously in \cite{Azeyanagi:2009pr}, with the result $\zeta/ \eta = 1/4 $.  

\section{Running bulk viscosity} \label{secRunning}

So far we have discussed solutions where the ratio between bulk and shear viscosity is a pure number, even for charged solutions. This makes the solutions \cite{Tarrio:2011de,Alishahiha:2012qu} quite special, we do not expect this to be true in general. 

Here we present an example of a charged solution (with $z=1$ and $\theta\neq 0$) where the bulk over shear viscosity ratio exhibits a non-trivial dependence with the temperature. This happens in the so-called `$ \gamma \delta =1$ solution', that was found in \cite{Charmousis:2009xr} and studied in detail in \cite{Charmousis:2010zz}. 
It is not straightforward to transform to the domain wall coordinate, thus we use the original presentation of the solution.

The action in this model is \eqref{holoaction}, with a single massless vector field and a scalar.
The scalar potential and its coupling to the vector field are parameterized by $\delta$
\begin{equation}
Z(\phi)=e^{\phi/\delta}, \ \qquad  V(\phi)=-2\Lambda e^{-\delta \phi}.
\end{equation}
This action admits a family of charged black hole solutions 
\begin{align}
	d s^2 &= - \frac{\mathcal V(r)d t^2 }{\mathcal F(r)^{c_0}}
	+ e^{\delta \phi}\frac{d r^2}{\mathcal V(r)} + r^2 \mathcal F(r)^{c_1}\Big(d x^2+d y^2\Big)\;,  \\
	e^{\phi}&= r^{2\delta } \mathcal F(r)^{c_2}\,, \\
	\mathcal A &= \frac{q}{(3-\delta ^2)r_+^{3-\delta ^2}}   \left(1
	- \left( \frac{r_+}{r}\right) ^{3-\delta ^2} \right)d t\,, \\
	\mathcal V(r) &= r^2-2 m ~r^{\delta ^2-1} + c_3 q^2 r^{2\delta ^2-4} \;,  \\
	\mathcal F(r) &= 1-\left(\frac{r_-}{r}\right)^{3-\delta ^2} \;, \\
	-\Lambda &= 3-\delta ^2\;, 
\end{align}
where 
\begin{align}
	&c_0 = \frac{4(1-\delta ^2)}{(3-\delta ^2)(1+\delta ^2)} \;, \\
	&c_1 = \frac{2(\delta ^2-1)^2}{(3-\delta ^2)(1+\delta ^2)} \;, \\
	&c_2 = \frac{4\delta (\delta ^2-1)}{(3-\delta ^2)(1+\delta ^2)} \;, \\
	&c_3 = \frac{(1+\delta ^2)}{4\delta ^2(3-\delta ^2)^2}  \;. 
\end{align}
$r_{\pm}$ are two roots of $\mathcal V(r)=0$ and give by 
\begin{align}
	&\left(r_{\pm}\right)^{3-\delta ^2} = m \pm \sqrt{m^2- c_3 q^2}\;. 
\end{align}
The black brane horizon sits at $r=r_+$ and there is curvature singularity at $ r=r_-$, beyond which the spacetime does not extend. 

Close to the boundary $r\to\infty$, the metric has the asymptotic form
	\begin{align} \label{bounmet}
		&d s^2 =r^{2} \left(- d t^2 + d x^2 + d y^2 \right) + r^{2\delta^2 -2} d r^2   \;.
	 \end{align}
Therefore, dual field theory has dynamical exponent $z=1$ and hyperscaling violation exponent $\theta = -\frac{d^2 \delta^2}{2 - d\delta^2}$ for $d=2$. Thus $ \delta^2 = -\frac{\theta}{2-\theta}$.

The parameters $m$ and $q$ are integration constants, which determine the gravitational mass and charge of the solution. In terms of these parameters, the 
temperature of the solution is given by 
\begin{align}
\notag T &= \frac{3-\delta^2}{4\pi} r_+^{1-\delta^2} \left(1 - \left(r_- / r_+\right)^{3-\delta^2} \right)^{1-c_1} \\
&= \frac{m^{1 - \frac{2}{3- \delta ^2}}2^{-\frac{5-\delta ^2}{3 -\delta ^2}}}{\pi} \left[ \left(3-\delta ^2\right) \!+\! \frac{\left(2 \delta ^4-3 \delta ^2-1\right) }{8 \delta ^2 \left(3-\delta ^2\right)^2}  \left( \frac{q}{m}\right)^2 \!+\! \cdots \! \right]. \label{temp} 
\end{align}
We expand the temperature for small charges for later use. 

Using the expressions for $r_{\pm}$, we can explicitly rewrite the entropy and scalar fields as $\phi^H = \phi^H(m, q)$ and $s=s(m, q) $.
\begin{align}
&s(m, q) =2^{c_1} Y^{c_1} \left(m+Y\right)^{-c_1-\frac{2}{\delta ^2-3}} \;, \\
&e^{\phi^H(m, q)} = 2^{c_2 } Y^{c_2} \left(m +Y\right)^{- c_2 -\frac{2 \delta }{\delta ^2-3}} \;, 
\end{align}
where $Y=\sqrt{m^2- c_3  q^2} $. 
Thus,  
\begin{align}
&d \phi^H =  \frac{\partial \phi^H}{\partial m} dm + \frac{\partial \phi^H}{\partial q} d q   \;, \\
&d s = \frac{\partial s}{\partial m} dm + \frac{\partial s}{\partial q} d q   \;. 
\end{align}
For the variation with fixed $q$, we get
	\begin{align} 
		s \frac{d\phi^H}{ds} &=s \frac{d \phi^H/d m}{d s/d m}  
		=\frac{c_2  m~ Y \left(\delta ^2-3\right) 	-Y^2 ( 2\delta + c_2  \left(\delta ^2-3\right) )}
		{c_1  m ~Y \left(\delta ^2-3\right) 
		- Y^2  \left(2+ c_1  \left(\delta ^2-3\right)\right)}  \;. 
	\end{align}
For the other contribution, we require $ ds = 0$ which gives $ dm =- \frac{ \partial s /\partial q}{\partial s /\partial m } d q $. 
	\begin{align} 
		q \frac{d\phi^H}{dq} & = \frac{\partial \phi^H}{\partial q} - \frac{\partial \phi^H}{\partial m} \frac{ \partial s /\partial q}{\partial s /\partial m } 
		=  \frac{2 c_3  q^2 (c_2 - c_1  \delta )}{c_3  q^2 \left(2 + c_1 \left(\delta ^2-3\right)\right)-2 m \left(m +Y\right)} \;. 
	\end{align}
Putting all together, the bulk viscosity becomes 
	\begin{align} 
		\frac{\zeta}{\eta} &= \left( s \frac{d\phi_i^H}{ds} + q \frac{d\phi_i^H}{d q}  \right)^2 
		= \frac{\left(1+\delta ^2\right)^2 X}{\left(\delta ^2-3\right)^2 \left(2\left(\delta ^2-1\right)^2+ \delta \sqrt{X} \right)^2} \;,
	\end{align}
where 
\begin{align}
X = 4  \delta ^2 \left(\delta ^2-3\right)^2- \left(\frac{q}{m}\right)^2 \left(1+\delta ^2\right) \;.
\end{align}

In order to see more clearly the effect of the charge on the bulk viscosity, we expand it for small charge $q/m\ll 1$
	\begin{align} 
		\frac{\zeta}{\eta} &= \delta ^2-\frac{\left(\delta ^2-1\right)^2 }{4  \left(\delta ^2-3\right)^2} \left(\frac{q}{m} \right)^2+\frac{3 \left(\delta ^2-1\right)^2 }{64 \left(\delta ^2-3\right)^3} \left(\frac{q}{m} \right)^4 + \cdots \;. 
	\end{align}
The leading term corresponds to the bulk viscosity in a neutral solution 
\begin{align} 
		\frac{\zeta_{0}}{\eta} &\sim \delta^2	= -\frac{2 \theta  }{d(d-\theta)} = -\frac{\theta  }{2-\theta}  \;.
	\end{align}
We compute the next to leading order correction as a function of temperature using \eqref{temp} 
 \begin{align} 
		\frac{\zeta_{1}}{\eta} &\sim -\frac{1}{4 (3-\theta )^2} \left(\frac{q}{m} \right)^2 
		\approx -\frac{ 4^{\theta-3 }}{ \pi ^{6-2 \theta }}  \frac{(3-\theta)^{4-2 \theta } }{(2-\theta)^{6-2 \theta } } ~
		\frac{q^2}{ T^{6-2 \theta}} \;.
	\end{align}

Combining these two contributions, the bulk viscosity reads 
	\begin{align} 
		\frac{\zeta}{\eta} &\approx -\frac{\theta  }{2-\theta} 
		-\frac{ 4^{\theta-3 }}{ \pi ^{6-2 \theta }}  \frac{(3-\theta)^{4-2 \theta } }{(2-\theta)^{6-2 \theta } } ~\frac{q^2}{ T^{6-2 \theta}} + \cdots \;.
	\end{align}
The interpretation of this formula is clear: at very high temperatures the properties of the system are determined by the UV physics, whose scaling properties are those of the metric \eqref{bounmet}, $z=1$ and $\theta\neq 0$. As we lower the temperature, the value of the bulk viscosity `runs' to a different value. At very low temperatures the charge is close to its critical value
\begin{equation}
q^2=m^2\left(\frac{1}{c_3}-\frac{\epsilon}{2}\right).
\end{equation} 
The temperature is in this case, to leading order in $\epsilon\ll 1$
\begin{equation}
T\simeq \frac{3-\delta^2}{4\pi}(c_3 q^2)^{(1-\delta^2)/2}\epsilon^{1-c_1}.
\end{equation}
Then,
\begin{equation}
\frac{\zeta}{\eta} \propto \left(\frac{T}{q^{1-\delta^2}}\right)^{\frac{2}{1-c_1}} =\left(\frac{T^{2-\theta}}{q^{2}}\right)^{\frac{(2-\theta)^3}{2(\theta^2-4\theta+1)}}
\end{equation}
Therefore, at very low temperatures the theory is `quasi-conformal', the value of the bulk viscosity is much smaller than the shear viscosity. This suggests that the IR theory possesses some kind of scale invariance.

\section{Outlook : Bulk viscosity on Quantum critical fluid} \label{secConclusion}

A hydrodynamic description for theories with Lifshitz scaling symmetry $z \neq 1$ has been put forward recently in \cite{Hoyos:2013eza,Hoyos:2013qna} as an effective description of quantum critical points \cite{SachdevBook,QCP,Hornreich:1975}. 
In particular, an analysis of the local entropy current along with the Landau frame condition reveals new transport coefficients contributing to the neutral and charged fluids at the first viscous order. These effects are direct consequence of the absence of boost invariance (Lorentz as well as Galilean boost). While the description is primarily oriented to condensed matter applications, relativistic Lorentz invariant models with broken boost invariance would have these effects, which are expected to be small, yet ubiquitous. 

Our general conclusion in the current paper is that for theories with a holographic dual the bulk viscosity vanishes unless the scaling symmetry is broken in some way. If this happens, it is sensitive to the details of the particular model, in particular it can depend on the charges. For the `neutral' cases, one can pin down the physical parameters $z$ and $\theta$ from the bulk viscosity and the speed of sound.

The dependence of thermodynamic quantities on the temperature in 
Lifshitz systems with dynamical exponent $z$ and hyperscaling violation exponent $\theta$ is
\begin{align}
s \sim T^{\frac{d-\theta}{z}} \;, \quad p \sim \frac{z}{z+d-\theta} T^{\frac{d+z-\theta}{z}} \;, \quad 
\epsilon \sim \frac{d-\theta}{z+d-\theta} T^{\frac{d+z-\theta}{z}} \;.
\end{align}
The speed of sound  is then
\begin{align}
c_s^2  &= \frac{\partial p}{\partial \epsilon} = \frac{z}{d-\theta} \;.
\end{align}
Taking also into account the bulk viscosity formula \eqref{BulkViscosity}, we have the following possibilities 
\begin{itemize}
\item Scale-invariant neutral systems constructed with massive vectors \cite{Balasubramanian:2009rx,Korovin:2013nha,Bertoldi:2009vn,Bertoldi:2009dt} : the bulk viscosity  vanishes and the speed of sound determines the dynamical exponent 
\begin{align}
\theta=0 \;, \quad z = d ~c_s^2 \;.
\end{align}
\item `Neutral' fluids with broken scale invariance constructed with scalar and massless vectors \cite{Charmousis:2009xr,Pang:2009ad,Gubser:2009qt,Goldstein:2009cv,Goldstein:2010aw,Charmousis:2010zz,Iizuka:2011hg,Gouteraux:2011ce,Cadoni:2011nq,Tarrio:2011de,Alishahiha:2012qu} : measuring bulk viscosity and speed of sound gives 
\begin{align}
c_s^2  &= \frac{\partial p}{\partial \epsilon} = \frac{z}{d-\theta}  \;, \qquad 
\frac{\zeta}{\eta} - c_s^2  = -2 \frac{d+\theta}{d(d-\theta)} \;. 
\end{align}

\end{itemize}

It would be interesting to see how these properties realized in the real physical materials, such as heavy fermion and high $T_c$ cuprates superconductors.

\section*{Acknowledgements}

We thank B. Gout\'eraux and E. Kiritsis for discussions and correspondences.  
This work is supported in part by the Israeli Science Foundation Center
of Excellence, and by the I-CORE program of Planning and Budgeting Committee and the Israel Science Foundation (grant number 1937/12). BSK is grateful for kind hospitality during the visit to Crete Center for Theoretical Physics, Heraklion.

\appendix


\section{Einstein and Maxwell equations} \label{EinsteinEQ}

In this appendix, we list the Einstein and Maxwell equations for the action \eqref{holoaction}. 
For simplicity, we consider the case with a scalar and a massive vector. For massless vector, one can set $m=0$. 
For rotationally invariant backgrounds, the Einstein equations are 
\begin{align}
0 &= 2 d A''  + \phi'^2  + e^{-2 ( A + g)} m^2 A_t^2\,, \label{AppEE} \\
0 &= g'' + g'^2 + (d+1)A' g' -e^{-2 (A+g) } m^2 A_t^2-e^{-2 A-g} Z(\phi) A_t'^2 \,,\\
0 &= 2 d A'(g'+(d+1)A') - 2 e^{-g} V(\phi) - \phi'^2  -e^{-2 (A+g) } m^2 A_t^2 +e^{-2 A-g} A_t'^2\;, 
\end{align} 
where $d$ is the number of spatial dimensions. 

The Maxwell equation is 
\begin{align} \label{AppMax}
&0 = - e^{(d-1) A-g} m^2 V_t + \left(e^{(d-1) A} Z(\phi) V_t' \right)'  \;. 
\end{align} 
There are two different cases. For massless vector field ($m=0$), we can simply get 
\begin{align}
e^{(d-1) A} Z(\phi) V_t' = Q \;.
\end{align}
Many available analytical solutions with Lifshitz asymptotics give 
\begin{align}
V_t \propto Q r^{\frac{d(d+z-\theta)}{(z-1)d - \theta}} f(r) \propto \rho^{\theta-d-z} f(\rho)\;.
\end{align}
$\rho$ coordinate gives more intuitive picture for scaling geometries and has been widely used 
\begin{align}
d s^2 =\rho^{-2 +2\theta/d} \left(- \rho^{-2(z-1)} f(\rho) d t^2 
		+ d x_i d x^i +\frac{d \rho^2}{f(\rho)}\right) \;.
\end{align}
Domain wall and $\rho$ coordinates are connected by a coordinate transformation $r \sim \rho^{1-z+\theta/d}$.

The solution of the massive vector field contrasts to that of the massless one. We consider only $\theta=0$.  
\begin{align}
V_t \propto r^{\frac{z}{z-1}} f(r) \propto \rho^{-z} f(\rho)\;.
\end{align}
For $\theta=0$ without the hyperscaling violation, we see that the massive vector has less divergent behavior at the boundary, and the massive vector provides a marginal deformation that preserve Lifshitz symmetry explained in detail in the main text. 

Scalar equation for all the cases has been checked to satisfy.


\begin{thebibliography}{aaaa}

\def\hri#1#2{{\color{blue}{\href{http://arxiv.org/abs/#1}{[#1]#2}}}}
\def\hre#1#2{{\color{blue}{\href{http://arxiv.org/abs/#1/#2}{[arXiv:#1/#2]}}}}

\vspace{-0.1in}

\bibitem{Baier:2007ix} 
  R.~Baier, P.~Romatschke, D.~T.~Son, A.~O.~Starinets and M.~A.~Stephanov,
  ``Relativistic viscous hydrodynamics, conformal invariance, and holography,''
  JHEP {\bf 0804}, 100 (2008)
  \hri{arXiv:0712.2451}{[hep-th]}.

\bibitem{Hoyos:2013eza} 
  C.~Hoyos, B.~S.~Kim and Y.~Oz,
  ``Lifshitz Hydrodynamics,''
  JHEP {\bf 1311}, 145 (2013)
  \hri{arXiv:1304.7481}{[hep-th]}.

\bibitem{Hoyos:2013qna} 
  C.~Hoyos, B.~S.~Kim and Y.~Oz,
  ``Lifshitz Field Theories at Non-Zero Temperature, Hydrodynamics and Gravity,''
  \hri{arXiv:1309.6794}{[hep-th]}.

\bibitem{Kachru:2008yh} 
  S.~Kachru, X.~Liu and M.~Mulligan,
  ``Gravity Duals of Lifshitz-like Fixed Points,''
  Phys.\ Rev.\ D {\bf 78}, 106005 (2008)
  \hri{arXiv:0808.1725}{[hep-th]}.

\bibitem{Taylor:2008tg} 
  M.~Taylor,
  ``Non-relativistic holography,''
  \hri{arXiv:0812.0530}{[hep-th]}.

\bibitem{Gath:2012pg} 
  J.~Gath, J.~Hartong, R.~Monteiro and N.~A.~Obers,
  ``Holographic Models for Theories with Hyperscaling Violation,''
  JHEP {\bf 1304}, 159 (2013)
  \hri{arXiv:1212.3263}{[hep-th]}.
  
\bibitem{Eling:2011ms} 
  C.~Eling and Y.~Oz,
  ``A Novel Formula for Bulk Viscosity from the Null Horizon Focusing Equation,''
  JHEP {\bf 1106}, 007 (2011)
  \hri{arXiv:1103.1657}{[hep-th]}.

\bibitem{Eling:2011ct}
  C.~Eling and Y.~Oz,
  ``Holographic Screens and Transport Coefficients in the Fluid/Gravity Correspondence,''
  Phys.\ Rev.\ Lett.\  {\bf 107} (2011) 201602
  \hri{arXiv:1107.2134}{[hep-th]}.

\bibitem{Eling:2009pb} 
  C.~Eling, I.~Fouxon and Y.~Oz,
  ``The Incompressible Navier-Stokes Equations From Membrane Dynamics,''
  Phys.\ Lett.\ B {\bf 680}, 496 (2009)
  \hri{arXiv:0905.3638}{[hep-th]}.

\bibitem{Eling:2009sj} 
  C.~Eling and Y.~Oz,
  ``Relativistic CFT Hydrodynamics from the Membrane Paradigm,''
  JHEP {\bf 1002}, 069 (2010)
  \hri{arXiv:0906.4999}{[hep-th]}.

\bibitem{Eling:2010hu}
  C.~Eling, Y.~Neiman and Y.~Oz,
  ``Holographic Non-Abelian Charged Hydrodynamics from the Dynamics of Null Horizons,''
  JHEP {\bf 1012} (2010) 086
  \hri{arXiv:1010.1290}{[hep-th]}.
  
\bibitem{Mas:2007ng} 
  J.~Mas and J.~Tarrio,
  ``Hydrodynamics from the Dp-brane,''
  JHEP {\bf 0705}, 036 (2007)
  \hre{hep-th}{0703093}.

\bibitem{Benincasa:2006ei} 
  P.~Benincasa and A.~Buchel,
  ``Hydrodynamics of Sakai-Sugimoto model in the quenched approximation,''
  Phys.\ Lett.\ B {\bf 640}, 108 (2006)
  \hre{hep-th}{0605076}.

\bibitem{Kanitscheider:2009as} 
  I.~Kanitscheider and K.~Skenderis,
  ``Universal hydrodynamics of non-conformal branes,''
  JHEP {\bf 0904}, 062 (2009)
  \hri{arXiv:0901.1487}{[hep-th]}.

\bibitem{Benincasa:2005iv} 
  P.~Benincasa, A.~Buchel and A.~O.~Starinets,
  ``Sound waves in strongly coupled non-conformal gauge theory plasma,''
  Nucl.\ Phys.\ B {\bf 733}, 160 (2006)
  \hre{hep-th}{0507026}.

\bibitem{Buchel:2008uu} 
  A.~Buchel and C.~Pagnutti,
  ``Bulk viscosity of N=2* plasma,''
  Nucl.\ Phys.\ B {\bf 816}, 62 (2009)
  \hri{arXiv:0812.3623}{[hep-th]}.

\bibitem{Buchel:2010gd} 
  A.~Buchel,
  ``Critical phenomena in N=4 SYM plasma,''
  Nucl.\ Phys.\ B {\bf 841}, 59 (2010)
  \hri{arXiv:1005.0819}{[hep-th]}.

\bibitem{Yarom:2009mw} 
  A.~Yarom,
  ``Notes on the bulk viscosity of holographic gauge theory plasmas,''
  JHEP {\bf 1004}, 024 (2010)
  \hri{arXiv:0912.2100}{[hep-th]}.

\bibitem{Gubser:2008sz} 
  S.~S.~Gubser, S.~S.~Pufu and F.~D.~Rocha,
  ``Bulk viscosity of strongly coupled plasmas with holographic duals,''
  JHEP {\bf 0808}, 085 (2008)
  \hri{arXiv:0806.0407}{[hep-th]}.

\bibitem{Gursoy:2009kk} 
  U.~Gursoy, E.~Kiritsis, G.~Michalogiorgakis and F.~Nitti,
  ``Thermal Transport and Drag Force in Improved Holographic QCD,''
  JHEP {\bf 0912}, 056 (2009)
  \hri{arXiv:0906.1890}{[hep-ph]}.

\bibitem{Balasubramanian:2009rx} 
  K.~Balasubramanian and J.~McGreevy,
  ``An Analytic Lifshitz black hole,''
  Phys.\ Rev.\ D {\bf 80}, 104039 (2009)
  \hri{arXiv:0909.0263}{[hep-th]}.

\bibitem{Korovin:2013nha} 
  Y.~Korovin, K.~Skenderis and M.~Taylor,
  ``Lifshitz from AdS at finite temperature and top down models,''
  JHEP {\bf 1311}, 127 (2013)
  \hri{arXiv:1306.3344}{[hep-th]}.

\bibitem{Bertoldi:2009vn} 
  G.~Bertoldi, B.~A.~Burrington and A.~Peet,
  ``Black Holes in asymptotically Lifshitz spacetimes with arbitrary critical exponent,''
  Phys.\ Rev.\ D {\bf 80}, 126003 (2009)
  \hri{arXiv:0905.3183}{[hep-th]}.

\bibitem{Bertoldi:2009dt} 
  G.~Bertoldi, B.~A.~Burrington and A.~W.~Peet,
  ``Thermodynamics of black branes in asymptotically Lifshitz spacetimes,''
  Phys.\ Rev.\ D {\bf 80}, 126004 (2009)
  \hri{arXiv:0907.4755}{[hep-th]}.

\bibitem{Kovtun:2004de} 
  P.~Kovtun, D.~T.~Son and A.~O.~Starinets,
  ``Viscosity in strongly interacting quantum field theories from black hole physics,''
  Phys.\ Rev.\ Lett.\  {\bf 94}, 111601 (2005)
  \hre{hep-th}{0405231}.

\bibitem{Charmousis:2009xr} 
  C.~Charmousis, B.~Gouteraux and J.~Soda,
  ``Einstein-Maxwell-Dilaton theories with a Liouville potential,''
  Phys.\ Rev.\ D {\bf 80}, 024028 (2009)
  \hri{arXiv:0905.3337}{[gr-qc]}.

\bibitem{Pang:2009ad} 
  D.~-W.~Pang,
  ``A Note on Black Holes in Asymptotically Lifshitz Spacetime,''
  \hri{arXiv:0905.2678}{[hep-th]}.

\bibitem{Gubser:2009qt} 
  S.~S.~Gubser and F.~D.~Rocha,
  ``Peculiar properties of a charged dilatonic black hole in AdS$_5$,''
  Phys.\ Rev.\ D {\bf 81}, 046001 (2010)
  \hri{arXiv:0911.2898}{[hep-th]}.

\bibitem{Goldstein:2009cv} 
  K.~Goldstein, S.~Kachru, S.~Prakash and S.~P.~Trivedi,
  ``Holography of Charged Dilaton Black Holes,''
  JHEP {\bf 1008}, 078 (2010)
  \hri{arXiv:0911.3586}{[hep-th]}.
  
\bibitem{Goldstein:2010aw} 
  K.~Goldstein, N.~Iizuka, S.~Kachru, S.~Prakash, S.~P.~Trivedi and A.~Westphal,
  ``Holography of Dyonic Dilaton Black Branes,''
  JHEP {\bf 1010}, 027 (2010)
  \hri{arXiv:1007.2490}{[hep-th]}.

\bibitem{Charmousis:2010zz} 
  C.~Charmousis, B.~Gouteraux, B.~S.~Kim, E.~Kiritsis and R.~Meyer,
  ``Effective Holographic Theories for low-temperature condensed matter systems,''
  JHEP {\bf 1011}, 151 (2010)
  \hri{arXiv:1005.4690}{[hep-th]}.
  
\bibitem{Iizuka:2011hg} 
  N.~Iizuka, N.~Kundu, P.~Narayan and S.~P.~Trivedi,
  ``Holographic Fermi and Non-Fermi Liquids with Transitions in Dilaton Gravity,''
  JHEP {\bf 1201}, 094 (2012)
  \hri{arXiv:1105.1162}{[hep-th]}.

\bibitem{Gouteraux:2011ce} 
  B.~Gouteraux and E.~Kiritsis,
  ``Generalized Holographic Quantum Criticality at Finite Density,''
  JHEP {\bf 1112}, 036 (2011)
  \hri{arXiv:1107.2116}{[hep-th]}.

\bibitem{Cadoni:2011nq} 
  M.~Cadoni, S.~Mignemi and M.~Serra,
  ``Exact solutions with AdS asymptotics of Einstein and Einstein-Maxwell gravity minimally coupled to a scalar field,''
  Phys.\ Rev.\ D {\bf 84}, 084046 (2011)
  \hri{rXiv:1107.5979}{[gr-qc]}.

\bibitem{Tarrio:2011de} 
  J.~Tarrio and S.~Vandoren,
  ``Black holes and black branes in Lifshitz spacetimes,''
  JHEP {\bf 1109}, 017 (2011)
  \hri{arXiv:1105.6335}{[hep-th]}.

\bibitem{Alishahiha:2012qu} 
  M.~Alishahiha, E.~O Colgain and H.~Yavartanoo,
  ``Charged Black Branes with Hyperscaling Violating Factor,''
  JHEP {\bf 1211}, 137 (2012)
  \hri{arXiv:1209.3946}{[hep-th]}.

\bibitem{Danielsson:2009gi} 
  U.~H.~Danielsson and L.~Thorlacius,
  ``Black holes in asymptotically Lifshitz spacetime,''
  JHEP {\bf 0903}, 070 (2009)
  \hri{arXiv:0812.5088}{[hep-th]}.

\bibitem{Mann:2009yx} 
  R.~B.~Mann,
  ``Lifshitz Topological Black Holes,''
  JHEP {\bf 0906}, 075 (2009)
  \hri{arXiv:0905.1136}{[hep-th]}.

\bibitem{Brynjolfsson:2009ct} 
  E.~J.~Brynjolfsson, U.~H.~Danielsson, L.~Thorlacius and T.~Zingg,
  ``Holographic Superconductors with Lifshitz Scaling,''
  J.\ Phys.\ A {\bf 43}, 065401 (2010)
  \hri{arXiv:0908.2611}{[hep-th]}.

\bibitem{Pang:2009pd} 
  D.~-W.~Pang,
  ``On Charged Lifshitz Black Holes,''
  JHEP {\bf 1001}, 116 (2010)
  \hri{arXiv:0911.2777}{[hep-th]}.

\bibitem{Dehghani:2011tx} 
  M.~H.~Dehghani, R.~B.~Mann and R.~Pourhasan,
  ``Charged Lifshitz Black Holes,''
  Phys.\ Rev.\ D {\bf 84}, 046002 (2011)
  \hri{arXiv:1102.0578}{[hep-th]}.

\bibitem{Barclay:2012he} 
  L.~Barclay, R.~Gregory, S.~Parameswaran, G.~Tasinato and I.~Zavala,
  ``Lifshitz black holes in IIA supergravity,''
  JHEP {\bf 1205}, 122 (2012)
  \hri{arXiv:1203.0576}{[hep-th]}.

\bibitem{Gouteraux:2011qh} 
  B.~Gouteraux, J.~Smolic, M.~Smolic, K.~Skenderis and M.~Taylor,
  ``Holography for Einstein-Maxwell-dilaton theories from generalized dimensional reduction,''
  JHEP {\bf 1201}, 089 (2012)
  \hri{arXiv:1110.2320}{[hep-th]}.

\bibitem{Cadoni:2009xm} 
  M.~Cadoni, G.~D'Appollonio and P.~Pani,
  ``Phase transitions between Reissner-Nordstrom and dilatonic black holes in 4D AdS spacetime,''
  JHEP {\bf 1003}, 100 (2010)
  \hri{arXiv:0912.3520}{[hep-th]}.

\bibitem{Bertoldi:2010ca} 
  G.~Bertoldi, B.~A.~Burrington and A.~W.~Peet,
  ``Thermal behavior of charged dilatonic black branes in AdS and UV completions of Lifshitz-like geometries,''
  Phys.\ Rev.\ D {\bf 82}, 106013 (2010)
  \hri{arXiv:1007.1464}{[hep-th]}.
  
\bibitem{Bertoldi:2011zr} 
  G.~Bertoldi, B.~A.~Burrington, A.~W.~Peet and I.~G.~Zadeh,
  ``Lifshitz-like black brane thermodynamics in higher dimensions,''
  Phys.\ Rev.\ D {\bf 83}, 126006 (2011)
  \hri{arXiv:1101.1980}{[hep-th]}.

\bibitem{Chemissany:2011mb} 
  W.~Chemissany and J.~Hartong,
  ``From D3-Branes to Lifshitz Space-Times,''
  Class.\ Quant.\ Grav.\  {\bf 28}, 195011 (2011)
  \hri{arXiv:1105.0612}{[hep-th]}.

\bibitem{Berglund:2011cp} 
  P.~Berglund, J.~Bhattacharyya and D.~Mattingly,
  ``Charged Dilatonic AdS Black Branes in Arbitrary Dimensions,''
  JHEP {\bf 1208}, 042 (2012)
  \hri{arXiv:1107.3096}{[hep-th]}.

\bibitem{Ogawa:2011bz} 
  N.~Ogawa, T.~Takayanagi and T.~Ugajin,
  ``Holographic Fermi Surfaces and Entanglement Entropy,''
  JHEP {\bf 1201}, 125 (2012)
  \hri{arXiv:1111.1023}{[hep-th]}.

\bibitem{Huijse:2011ef} 
  L.~Huijse, S.~Sachdev and B.~Swingle,
  ``Hidden Fermi surfaces in compressible states of gauge-gravity duality,''
  Phys.\ Rev.\ B {\bf 85}, 035121 (2012)
  \hri{arXiv:1112.0573}{[cond-mat.str-el]}.

\bibitem{Azeyanagi:2009pr} 
  T.~Azeyanagi, W.~Li and T.~Takayanagi,
  ``On String Theory Duals of Lifshitz-like Fixed Points,''
  JHEP {\bf 0906}, 084 (2009)
  \hri{arXiv:0905.0688}{[hep-th]}.

\bibitem{SachdevBook} 
S.~Sachdev, 
{\em "Quantum Phase Transitions,"}
2nd Ed., Cambridge University Press (2011).

\bibitem{QCP}
P.~{Gegenwart}, Q.~{Si}, and F.~{Steglich}, {\it {Quantum criticality in
  heavy-fermion metals}},  {\em Nature Physics} {\bf 4}, 186 (2008)
  \hri{arXiv:0712.2045}{[cond-mat.str-el]}.

\bibitem{Hornreich:1975}
R.~M. Hornreich, M.~Luban, and S.~Shtrikman, {\it Critical behavior at the
  onset of $\vec{k}$-space instability on the $\lambda$ line},  {\em Phys. Rev.
  Lett.} {\bf 35}, 1678 (1975).

\end{thebibliography}
\end{document}